# INVESTIGATION ON THE INITIAL SYSTEM ANGLE OF A TORSION SPRINGS USING CAD

Paredes Manuel[*]

Institut Clément Ader (ICA), Université de Toulouse, INSA, CNRS FRE 3687, France

*manuel.paredes@insa-toulouse.fr

**RESUMEN**

Helical torsion springs with tangential legs are used for many applications: from simple everyday use systems, like clothes peg, to advanced systems, as sectional doors. Torsion springs are usually exploited with an inner rod as a guide. The required space between the spring and the rod make the spring tilted. Unfortunately, as far as we are aware, no industrial software is able to determine the initial system angle made by the spring in its tilted position. All the torque/angle curves are presented with relative angles. It means that whatever the number of coils considered, the angle is null when the torque is null. For that reason, we have developed a methodology using both CATIA and ABAQUS to determine the system angle at the beginning of the behavior. The model exploits wires designed with CATIA and imported in ABAQUS for the center rod, the torsion spring and the rods of the system to apply the torque. Several options to model contact between the several parts have been investigated. They mostly use connector with an axial property and a stop criterion. The obtained results are very interesting and enable to plan further studies.

# 1. Context of the study

Mechanical compression springs are often used in mechanical devices for their ability to store and return energy. The range of applications is very wide. When rotating parts are involved helical torsion springs can be exploited to define the relation between torque and rotating angle.

In the preliminary design phase, the use is commonly to first define the parts geometries and then to find a suitable torsion spring that will give the expected torques for at least 2 system angles (the two bounds of the travel). An example modeled with Catia is illustrated in Figure 1. It shows the inner rode which defines the rotating axle and the two rods that will be in contact with the legs of the torsion spring. The system angle [AOB] is highlighted.

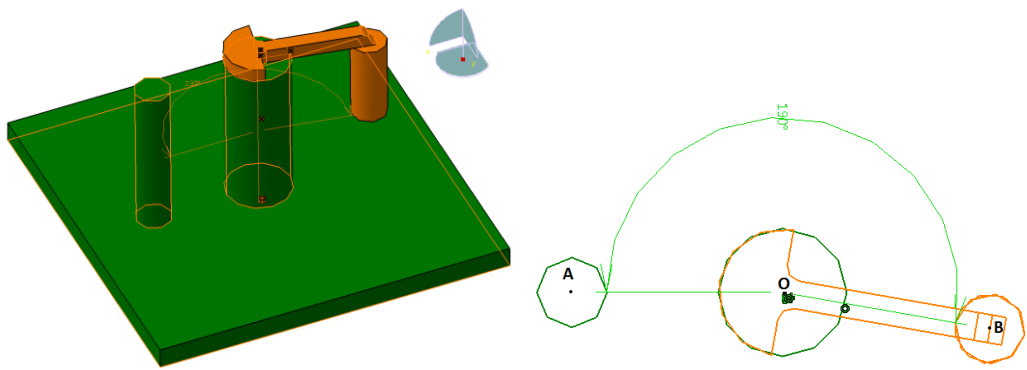

Figure 1. Example of rotating system

The system been designed, the next step is to find a suitable spring for the application. To help them, designers can exploit the book of Wahl [1] which is considered as the bible of spring design. They can also refers to standards [2] or directly exploit industrial software dedicated to spring design [3-4]. Unfortunately, all the torque VS angle relations are defined once the contact is established and none give an evaluation of the system angle. Designers can thus evaluate the rate of the spring but can't evaluate when the spring will start to work in the studied system.

For a given system and a given torsion spring, it is of key interest for a designer to evaluate the initial system angle. Moreover, the required space between the spring and the inner rod make the spring tilted. Evaluating the system angle is thus a non basic issue. As a first step to satisfy this need, this paper presents a study that exploits both CATIA and Abaqus to evaluate the initial system angle.

## 2. Using Catia and Abaqus to evaluate the system angle

To model spring, the use is to exploit 3D software as Catia to obtain the geometry and then to exploit 3D FEA in Catia itself or with dedicated software as Abaqus. To go further, it is necessary to consider how torsion springs are manufactured. Helical torsion springs are usually manufactured with contact between coils and no pretension. It induces that the axial pitch is equal to the wire diameter and that coils are tangential. This manufacturing issue unables to exploit the common process of 3D modeling and automatic meshing for FEA as all the coils of the spring will be merged at the tangential points. For that reason Shimoseki et al. [5] have proposed to use wires to model the spring and the system. An example with a spring with tangential legs and 9.25 coils is shown in figure 2.

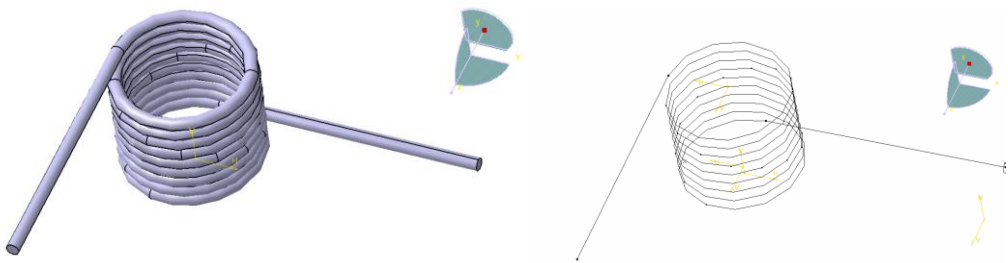

Figure 1. Example of a torsion spring with 3D and wire modelling

The wire associated to the torsion spring can be imported in FEA software. The next step is to simulate contact between the spring and the three rods and many options are available. As an illustration example, this paper presents a simulation that exploits 3 differents ways to model contact between cylinders that are modeled using wires.

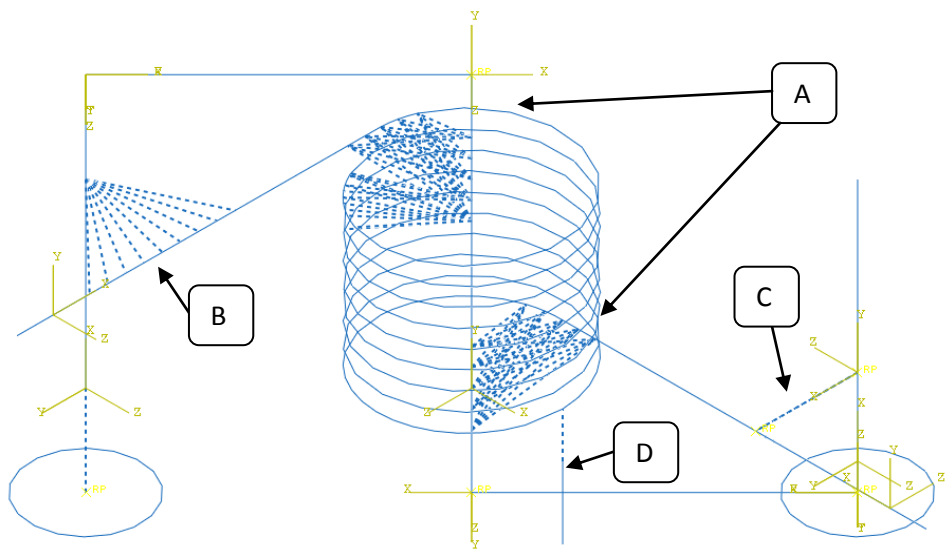

Figure 3. Modeling using Abaqus

Figure 3 higlights the model exploited with Abaqus. In Area A, the contact between the inner rod and the spring is modeled by axial connections between points. In area B, the contact between the spring and the rotating rod is modelled by a radial-thrust connection. In area C, 2 wire parts are added and 3 cylindrical connexions are exploited. Boundary conditions enable the rotating rod to move arround the inner rod axis only and tie the other rods to the ground. Finally a fexible beam is added to link the spring to the ground in area D and avoid inconsistency.

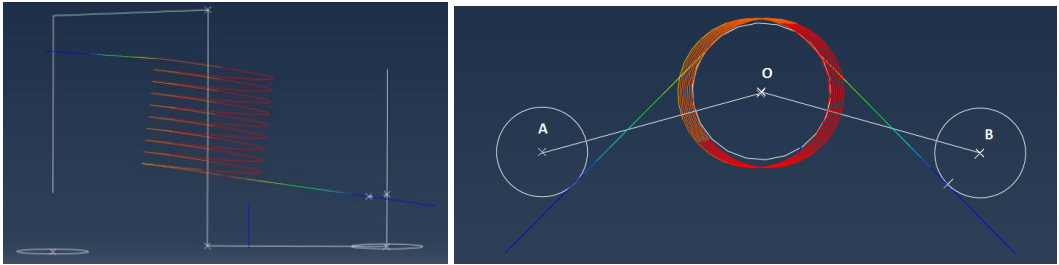

Figure 4. Results

The simulation (Figure 4 left) clearly shows that the spring is tilted when the contact is established. Moreover, the initial system angle [AOB] can be evaluated 210°, (figure 4 right) and differs from the angle between legs of the spring (270° for a spring with 9.25 coils).

## 3. Conclusions

We have shown that the acurate design of helical torsion spring requires the evaluation of the system angle when the spring starts to be in contact with the rods. Simulation software can help to satisfy this need. In this paper with have shown that using wire modeling can be of key interest. Moreover we have highlighted that several modeling options can be exploited. The wire modeling induces a fast solving process. Thus, further works may enable to automated both modeling and solving process. Many expermients may be simulated in order to identify analytical laws that could be directly used by designers in the preliminary design stages.